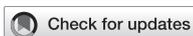

# Spatial-temporal evolution characteristics and driving factors of carbon emission prediction in China-research on ARIMA-BP neural network algorithm

Sanglin Zhao[1]*, Zhetong Li[2], Hao Deng[1], Xing You[3], Jiaang Tong[4]*, Bingkun Yuan[5]* and Zihao Zeng[1]

[1]School of Engineering Management, Hunan University of Finance and Economics, Changsha, China, [2]School of Mathematics and Statistics, Hunan University of Finance and Economics, Changsha, China, [3]School of Economics, Hunan University of Finance and Economics, Changsha, China, [4]School of Economics, Fudan University, Shanghai, China, [5]School of Energy Science and Engineering, Central South University, Changsha, China

China's total carbon emissions account for one-third of the world's total. How to reach the peak of carbon emissions by 2030 and achieve carbon neutrality by 2060 is an important policy orientation at present. Therefore, it is of great significance to analyze the characteristics and driving factors of temporal and spatial evolution on the basis of effective calculation and prediction of carbon emissions in various provinces for promoting high-quality economic development and realizing carbon emission reduction. Based on the energy consumption data of 30 provinces in China from 2000 to 2021, this paper calculates and predicts the total carbon emissions of 30 provinces in China from 2000 to 2035 based on ARIMA model and BP neural network model, and uses ArcGIS and standard elliptic difference to visually analyze the spatial and temporal evolution characteristics, and further uses LMDI model to decompose the driving factors affecting carbon emissions. The results show that: (1) From 2000 to 2035, China's total carbon emissions increased year by year, but the growth rate of carbon emissions gradually decreased; The carbon emission structure is "secondary industry > residents' life > tertiary industry > primary industry", and the growth rate of carbon in secondary industry and residents' life is faster, while the change trend of primary industry and tertiary industry is smaller; (2) The spatial distribution of carbon emissions in China's provinces presents a typical pattern of "eastern > central > western" and "northern > southern", and the carbon emission centers tend to move to the northwest; (3) The regions with high level of digital economy, advanced industrial structure and new quality productivity have relatively less carbon emissions, which has significant group difference effect; (4) The intensity effect of energy consumption is the main factor driving the continuous growth of carbon emissions, while the *per capita* GDP and the structure effect of energy consumption are the main factors restraining carbon emissions, while the effects of industrial structure and population size are relatively small. Based on the research conclusion, this paper puts forward some policy suggestions from energy structure, industrial structure, new quality productivity and digital economy.

# 1 Introduction

First of all, China plays a crucial role in reducing carbon emissions. Chinese government stated that "China would actively and steadily promote carbon neutrality in peak carbon dioxide emissions". As can be seen from Figure 1, China's total carbon emissions have been rising rapidly since 2000, and the total carbon emissions in 2023 will be 12.6 billion tons, accounting for one-third of the global carbon emissions of 37.4 billion tons, making it the largest carbon emitter in the world.[1]

Secondly, there are significant regional differences in carbon emissions. In 2023, the province with the highest carbon emission is Shanxi, and the lowest is Qinghai. East China is the largest carbon emission region, accounting for about 30% of the national proportion.[2] There are significant differences in carbon emissions among provinces, and it is an important way to achieve a balanced and coordinated carbon emission market.

Controlling the total amount of carbon emissions and optimizing the carbon emission structure are necessary methods to achieve the "carbon peaking and carbon neutrality goals". Therefore, based on the these problems, this paper combines ARIMA time series analysis and BP neural network algorithm to measure and predict the carbon emissions of the whole country and provinces, and analyzes the spatial evolution characteristics of carbon emissions in China, while decomposes the driving factors by LMDI (Logarithmic Mean Divisia Index). The research conclusions provide decision-making reference for making scientific and reasonable emission reduction policies in different regions of China based on their local conditions.

## 1.1 Research significance

(1) A more comprehensive understanding of the regional distribution characteristics and evolution trend of China's carbon emissions. This paper measures and forecasts the total amount of carbon emissions from 2000 to 2035, analyzes the historical change trajectory and future development trend of carbon emissions data in different regions and industries, depicts the spatial distribution characteristics of carbon emissions in China, and provides scientific basis for the government to formulate more accurate emission reduction policies.

(2) The specific driving factors of carbon emission change are clarified. Through LMDI model, the factors affecting the change of carbon emissions are decomposed into six factors: energy consumption structure, energy consumption intensity, *per capita* GDP effect, population size effect and industrial structure effect. This research also analyzed how each of the above factors affects carbon emissions.

(3) Provide reference for making emission reduction policies according to local conditions. A group analysis was conducted based on the differences in industries, industrial structures, digital economy and new productivity in different regions. Based on the research conclusions, it provides scientific basis for regions at different development stages to formulate more reasonable emission reduction policies based on local conditions.

## 1.2 Research status

### 1.2.1 Carbon emission measurement method

At present, there are mainly four mainstream carbon emission prediction models. One is STIRPAT model. Zhou et al. (2024) introduced factors such as population, economy and energy structure, and used STIRPAT model to conduct in-depth analysis on the influencing factors of carbon emissions in the industrial sector of Jiangsu Province, and predicted the situation of carbon peak. Lu et al. (2024) based on STIRPAT model, combined with ridge regression model, predicted the spatio-temporal pattern evolution of carbon emissions and carbon peak path in Anhui Province. The second is leap model. Leap model has the characteristics of flexible structure, less data demand and can reflect technological progress and energy efficiency. Cheng (2023) predicted the energy consumption structure, energy demand and carbon emission trend of Jilin Province Based on leap model. On the other hand, Chen et al. (2024) used leap model to predict the carbon emission scenarios of power grid enterprises and analyzed the emission reduction potential. The third is LSTM model. Wang (2005) predicted the future carbon emission trend based on the LSTM model by mining the time-dependent and cyclical laws in the historical carbon emission data. Hu et al. (2022) obtained the prediction results of carbon emission intensity based on the LSTM model and ARIMA model. The fourth is ARIMA model. Nan et al. (2023) combined the exponential smoothing model with ARIMA model to predict the trend of industrial carbon emissions. This method combines the long-term trend of time series and short-term random changes, so as to improve the accuracy of prediction. Jia et al. (2024) and others proposed a hybrid model combining ARIMA model, multi-layer perceptron (MLP) and grasshopper optimization algorithm (GOA) for water quality time series prediction. However, the prediction stability and robustness of the hybrid model under extreme weather conditions or sudden pollution events are not discussed in detail, and the water quality data may show stronger nonlinearity and nonstationarity in these cases. Ni et al. (2025) used the variational mode decomposition (VMD) of infrasonic signal and ARIMA model to predict debris flow. However, the impact of natural factors such as topography and geological conditions on the accuracy of debris flow prediction was not fully considered, which played an important role in the occurrence of debris flow. Xu et al. (2024) analyzed and predicted the atmospheric environmental quality of Hunan Province Based on ARIMA model in the Journal of sustainability. However, the performance of ARIMA model in long-time scale prediction and how to deal with the problem of outliers and missing values in atmospheric environmental quality data are not discussed in depth. STIRPAT, leap, LSTM and ARIMA, the four mainstream carbon emission prediction models, have their own characteristics. STIRPAT model is good at comprehensiveness and applicability, and can comprehensively analyze the population, economy, energy structure and other factors, but the data demand is high and the model structure is complex; Leap model has flexible structure, small data demand, and can reflect technological progress and changes in energy efficiency, but it is highly subjective and the accuracy of long-term prediction may be reduced; LSTM model performs well in processing time series data and prediction accuracy, but it needs a lot of historical data and high computational resources; Although ARIMA model has advantages in time series analysis and prediction stability, it has weak nonlinear data processing ability and high data preprocessing requirements. Therefore, it is particularly important to use the combined optimization model.

---

[1] Data source: International Energy Agency (https://www.iea.org/reports/co2-emissions-in-2023).

[2] The East China region has the highest carbon emissions in China, accounting for about 30% of China's carbon emissions. This is based on the "National Carbon Market Development Report (2024)" and the "China Energy Statistical Yearbook" released by the Chinese Ministry of Ecology and Environment (https://www.mee.gov.cn/ywdt/xwfb/202407/t20240722_1082192.shtml).

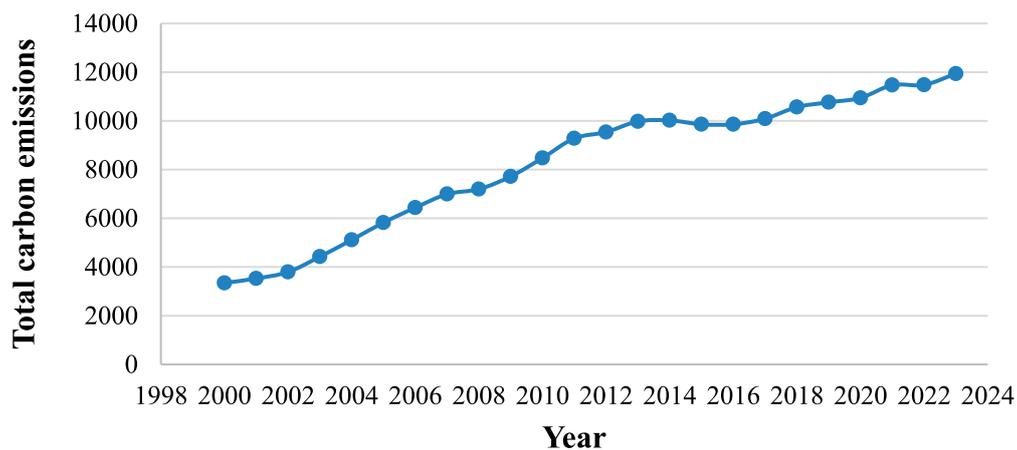

FIGURE 1
Total carbon emission of China from 2000 to 2023 (unit:million tons) Source: China Energy Statistics Yearbook. URL:https://www.stats.gov.cn/sj/ndsj.

### 1.2.2 Drivers of carbon emissions

The driving factors of carbon emission include energy consumption, energy technology, population and economic development. Liu (2018) calculated the carbon emissions of energy consumption in China in detail in his research, and concluded that energy intensity, energy structure and industrial structure are important factors affecting carbon emissions, and the reduction of energy intensity, the promotion of clean energy and the upgrading of industrial structure are the key ways to reduce carbon emissions. The research of Zhu et al. (2015) focuses on reducing carbon emission estimation in the process of fossil fuel combustion and cement production in China. By improving combustion technology and production process, carbon emission can be effectively reduced. Yuli et al. (2018) based on China's carbon emission accounts from 1997 to 2015, concluded that population size and economic development are important factors affecting carbon emissions. With the growth of population and the rapid development of economy, carbon emissions are increasing year by year. However, by improving energy efficiency and promoting clean energy, the growth rate of carbon emissions can be slowed down to some extent.

### 1.2.3 Literature review

The carbon emission measurement methods mentioned above, such as STIRPAT, LEAP, LSTM and ARIMA, all have limitations in dealing with complex nonlinear and time series deep models, and the driving factors of carbon emission do not put energy, population and economy into a unified framework analysis. Therefore, this paper constructs "ARIMA-BP neural network model", which not only has the advantages of ARIMA model in time series analysis, but also can deal with the nonlinear relationship through BP neural network model, improve the accuracy and reliability of prediction, capture the spatial and temporal characteristics of carbon emissions in China more comprehensively. Based on the LMDI model, the driving factors of the spatiotemporal characteristics of carbon emissions are

TABLE 1 Interpretation of carbon emission measurement variables.

| Variable name | Variable interpretation and assignment | Unit |
|---|---|---|
| EC | Energy carbon dioxide emissions | Ten thousand tons tCO2 |
| i | Energy consumption category | \ |
| Eij | The j-th energy consumption in the I-industry | Kilogram standard coal |
| rj | Carbon emission coefficient of the j-th energy source | tC/1012J |
| 44/12 | Conversion coefficient of carbon and carbon dioxide | \ |

Note: Due to the lack of data, the calculation of energy carbon emissions does not include the carbon emissions of electricity and heat in the consumer sector.

discovered, providing scientific support for carbon emission management and emission reduction policies.

# 2 Theoretical basis of carbon emission prediction

## 2.1 Definition of the concept of carbon emission measurement

Carbon emission refers to the greenhouse gas emission generated during the production, transportation, use and recycling of a product, and the carbon emission factor method is a commonly used method for measuring carbon emission. Based on the guidelines issued by official website, the Intergovernmental Panel on Climate Change of the United Nations, and the carbon emission factor method recommended by provincial units in China, this paper calculates the carbon emissions of energy consumption in three major industries (primary industry, secondary industry and tertiary industry). The specific formula is as follows:

$$EC = \sum_{i=1}^{3}\sum_{j=1}^{6} E_{ij} r_j \times \frac{44}{12} \quad (1)$$

The explanations of the variables in Formula 1 are shown in Table 1:

## 2.2 Theoretical model of carbon emission prediction

### 2.2.1 ARIMA model

ARIMA model is widely used in all kinds of time series data analysis and modeling. The total carbon emission is a typical time series data, and ARIMA model is used in this paper to make a preliminary prediction. The essence of ARIMA model is to transform non-stationary time series into stationary time series, and establish a model in which the dependent variable regresses the lag value and the current value and lag value of random error term. The specific formula is as follows:

$$y_t = \mu + \sum_{i=1}^{p} \gamma_i y_{t-1} + \varepsilon_t + \sum_{i=1}^{q} \theta_i \varepsilon_{t-i} \quad (2)$$

In Formula 2, the current value, the previous value, the constant term, the order, the autocorrelation coefficient, the error term, and q, the relationship between the current error and the previous errors, the coefficient. $y_t$ For the current value, $y_{t-1}$ Previous period value, $\mu$ As a constant term, $p$ For order, $\gamma$ For autocorrelation coefficients, $\varepsilon$ For error terms, $\theta$ For the coefficient.

### 2.2.2 BP neural network model

BP neural network is a multi-layer feedforward neural network, it's main feature is that the signal propagates forward and the error propagates backward. For the neural network model with two hidden layers, referring to the research of Hu et al. (2021), the process of BP neural network is mainly divided into two stages. The first stage is the forward propagation of signals, from the input layer through the hidden layer and finally to the output layer; The second stage is the reverse propagation of errors, from the output layer to the hidden layer, and finally to the input layer, the weights and offsets from the hidden layer to the output layer and from the input layer to the hidden layer are adjusted in turn.

### 2.2.3 LMDI carbon emission factor decomposition model

LMDI model decomposes carbon emissions into six factors: energy consumption structure, carbon emission coefficient, energy consumption intensity, industrial structure, *per capita* GDP and population size. The formula of carbon emission from energy consumption can be rewritten as follows:

$$C = \sum_{i=1}^{3}\sum_{j=1}^{6} \frac{E_{ij}}{E_i} \frac{C_{ij}}{E_{ij}} \frac{E_i}{G_i} \frac{G_i}{G} \times P$$

$$= \sum_{i=1}^{3}\sum_{j=1}^{6} s_{ij} \times f_{ij} \times e_i \times n_i \times r \times p \quad (3)$$

The variables in Formula 3 and their related explanations are shown in Table 2.

In order to study the sources and influencing factors of carbon emissions, the LMDI, decomposition formula is converted into a linear formula:

$$\Delta C = \Delta C_s + \Delta C_f + \Delta C_e + \Delta C_n + \Delta C_r + \Delta C_p \quad (4)$$

See Table 3 for the explanation of variables in Formula 4. The carbon emission factor remains basically unchanged, so there is no carbon emission effect, that is. $\Delta C_f = 0$.

Then we can measure the effect of each factor every year:

$$\Delta C_s = \sum_{i=1}^{3}\sum_{j=1}^{6} w(C_{ij}^{t-1}, C_{ij}^t) \ln \frac{s_{ij}(t)}{s_{ij}(t-1)} \quad (5)$$

TABLE 2 Explanation of variables in carbon emission formula of energy consumption.

| Variable name | Variable interpretation and assignment | Unit |
|---|---|---|
| P | population size | ten thousand people |
| G | GDP | hundred million yuan |
| $E_{ij}$ | The j-th energy consumption in the I-industry | Kilogram standard coal |
| $C_{ij}$ | Carbon Emissions of the J-type Energy in the I-industry | Ten thousand tons $CO_2$ |
| $s_{ij}$ | Consumption proportion of the j-th energy in the I-th industry | % |
| $f_{ij}$ | Carbon emission coefficient of the j-th energy in the I-th industry | \ |
| $e_i$ | Energy consumption intensity of the I industry | \ |
| $n_i$ | Proportion of GDP in the I industry | % |
| r | Per capita GDP | 100 million yuan/ten thousand people |

TABLE 3 Explanation of formula variables of LMDI decomposition model of carbon emissions.

| Variable name | Variable interpretation and assignment |
|---|---|
| $\Delta C$ | Comprehensive effect of incremental energy carbon emissions |
| $\Delta C_s$ | Structural effect of energy consumption |
| $\Delta C_f$ | Carbon emission coefficient effect |
| $\Delta C_e$ | Energy consumption intensity effect |
| $\Delta C_n$ | Industrial structure effect |
| $\Delta C_r$ | Per capita GDP effect |
| $\Delta C_p$ | Population scale effect |

$$\Delta C_e = \sum_{i=1}^{3}\sum_{j=1}^{6} w\left(C_{ij}^{t-1}, C_{ij}^{t}\right) \ln \frac{e_{ij}(t)}{e_{ij}(t-1)} \qquad (6)$$

$$\Delta C_n = \sum_{i=1}^{3}\sum_{j=1}^{6} w\left(C_{ij}^{t-1}, C_{ij}^{t}\right) \ln \frac{n_{ij}(t)}{n_{ij}(t-1)} \qquad (7)$$

$$\Delta C_r = \sum_{i=1}^{3}\sum_{j=1}^{6} w\left(C_{ij}^{t-1}, C_{ij}^{t}\right) \ln \frac{r_{ij}(t)}{r_{ij}(t-1)} \qquad (8)$$

$$\Delta C_p = \sum_{i=1}^{3}\sum_{j=1}^{6} w\left(C_{ij}^{t-1}, C_{ij}^{t}\right) \ln \frac{p_{ij}(t)}{p_{ij}(t-1)} \qquad (9)$$

Among them:

$$w\left(C_{ij}^{t-1}, C_{ij}^{t}\right) = \begin{cases} \frac{c_{ij}^{t} - c_{ij}^{t-1}}{\ln\left(c_{ij}^{t}/c_{ij}^{t-1}\right)}, C_{ij}^{t} \neq C_{ij}^{t-1} \\ C_{ij}^{t} \text{ or } C_{ij}^{t-1}, C_{ij}^{t} = C_{ij}^{t-1} \end{cases} \qquad (10)$$

# 3 Data and model construction

## 3.1 Variables and data sources

### 3.1.1 Source of data

The energy-related data studied in this paper mainly come from China Energy Statistical Yearbook, GDP, industry and population data come from China Statistical Yearbook and provincial statistical yearbooks, and the carbon emission factor comes from the specific parameters of the carbon emission calculation of the International Energy Agency. For some missing data, it is supplemented by manual collection on relevant government statistical websites. In order to avoid duplication of calculation, the processing conversion and energy consumption in the energy balance table are not included in the total calculation.

### 3.1.2 Description of variables

Firstly, the output value is classified into primary industry, secondary industry and tertiary industry.[3] Then we calculates the energy consumption per unit GDP of each industry. After multiplying the two, the corresponding energy consumption of each industry can be obtained. This paper takes the average value of the carbon emission factors in the previous 10 years as the carbon emission factor value for 2022-2035, multiplies the energy consumption of each industry by the carbon emission factor, and then multiplies it by the conversion coefficient and sums them up to get the total carbon emissions. Therefore, the variables involved in this paper mainly include the energy consumption of the three major industries. The output value of each industry, population and economic variables, and the descriptive statistics of the main variables are shown in Table 4.

## 3.2 Construction of ARIMA-BP neural network model

Firstly, the ARIMA model is used to predict the data, and the prediction error sequence is obtained. Secondly, a two-layer hidden layer BP neural network with one neuron and three neurons is established. Input the error data into the neural

---

3 The primary industry includes the agriculture and forestry consumption sector, the secondary industry includes the industrial consumption sector and the energy consumption sector, and the tertiary industry includes the construction consumption sector and the transportation consumption sector.

TABLE 4 Descriptive statistics of main variables.

| Variable name | Sample size | Maximum | Minimum value | Average value | Standard deviation |
|---|---|---|---|---|---|
| power | 3,284 | 5,600.89 | 0.39 | 253.72 | 568.89 |
| coal | 3,142 | 11,526.60 | 0 | 631.16 | 1,416.13 |
| heating power | 1738 | 142,256 | 0 | 5,607.78 | 11,401.62 |
| natural gas | 2017 | 165.13 | 0 | 10.47 | 18.05 |
| petroleum | 3,277 | 2,653 | 0.17 | 247.06 | 343.55 |
| Other energy sources | 778 | 1,556.51 | 0 | 79.72 | 131.59 |
| Output value of primary industry | 661 | 6,029 | 40.1 | 1,421.4 | 1,255.14 |
| Output value of secondary industry | 661 | 52,678.7 | 80.9 | 7,132.35 | 8,185.50 |
| Output value of tertiary industry | 661 | 69,179 | 127.6 | 8,084.38 | 9,974.76 |
| Regional GDP | 661 | 124,720 | 263.7 | 16,638.12 | 18,777.37 |

Note: As there are many energy sources, the six categories of energy sources are described and counted by combined data.

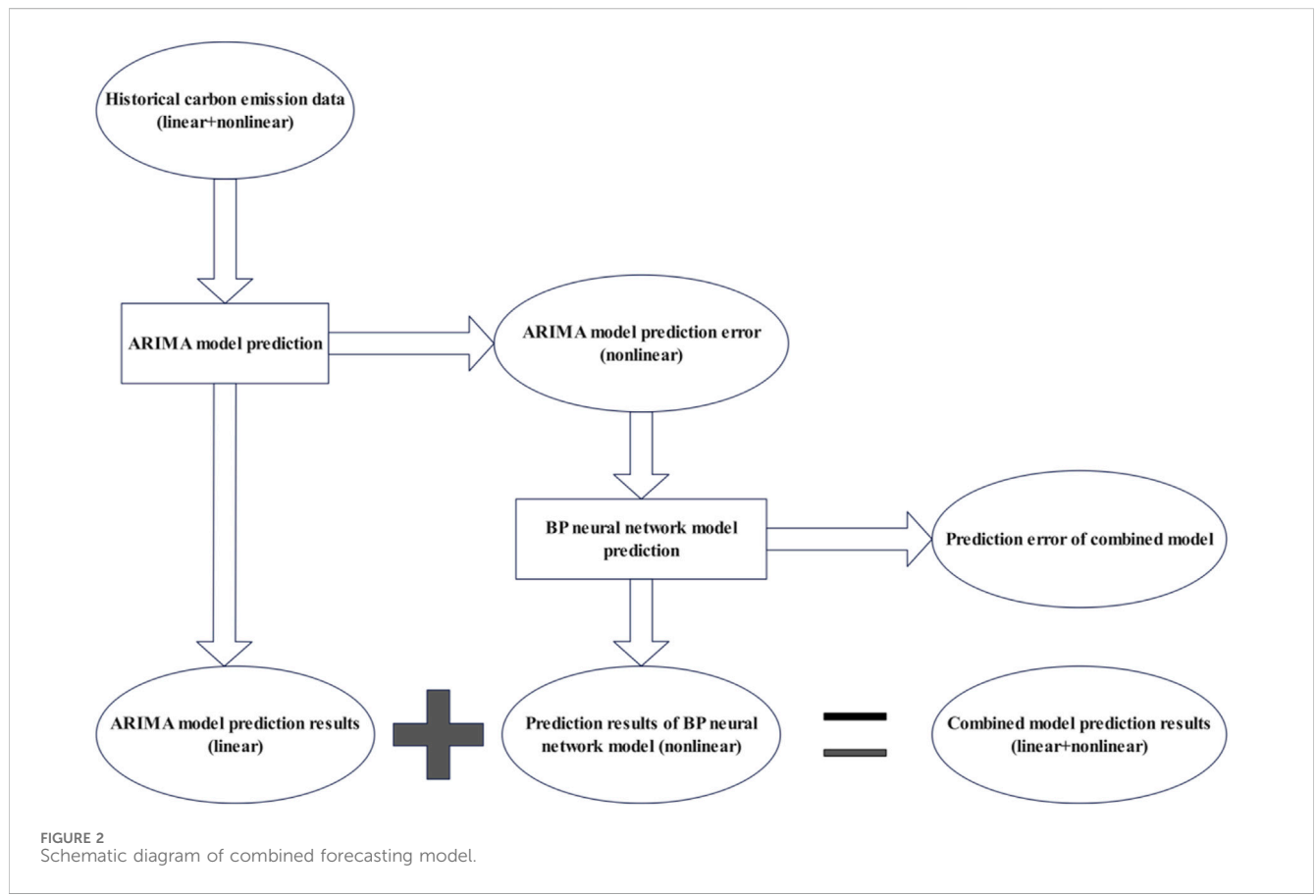

FIGURE 2
Schematic diagram of combined forecasting model.

network in time sequence, with the input node set to 4 and the output node set to 1. By rolling the window, the error of the previous period will continue to be transmitted into the neural network, and as part of the input, the model will be continuously revised and predicted. Finally, the combination model is used to predict the change of carbon emission intensity (Figures 2, 3). The specific steps are as follows:

### 3.2.1 ARIMA model construction

The construction of ARIMA model is divided into the following three processes:

(1) The stationarity test of time series. In this paper, the ADF method is used to test the stationarity of time series. The T statistic of adjoint probability is −1.86~-1.90, which can't reject the original hypothesis that there is a unit root at the significance level of 5%. The DCI sequence is obtained by taking the first-order difference of each sequence, and the ADF test is carried out again. The T statistic of adjoint probability is −2.97~-2.99, which rejects the hypothesis of unit root at the statistical level of 1%. After 1.

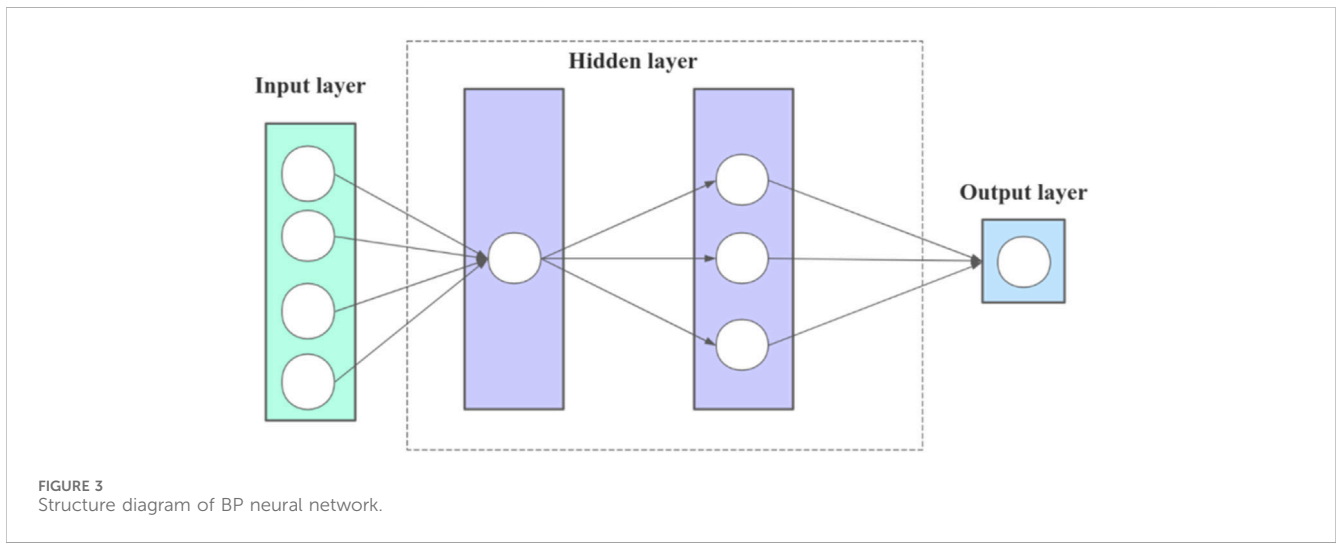

FIGURE 3
Structure diagram of BP neural network.

(2) Determine the order p of ARIMA model. In this paper, autocorrelation (AC) coefficient and partial autocorrelation (PAC) coefficient are adopted, and the order determination principle such as AIC is adopted. According to the comprehensive analysis of AC and PAC images, data and inspection table, (0,1,0), (0,1,1), (0,1,0), (0,2,1) and others model is adopted. The detailed reasons for model selection and ARIMA model verification diagram are shown in the annex and attached Tables 1, 2

(3) Parameter estimation and diagnostic test. The purpose of testing the significance of model parameters is to determine which parameters are effective in explaining the model. Validity test can help to determine whether the model can accurately predict or interpret data. At the same time, it is necessary to check whether the residual sequence is a white noise sequence to ensure that the residual of the model has no missing information or structure.

### 3.2.2 BP neural network model construction

Because the carbon emission intensity is influenced by economic factors and carbon emission activities, it is difficult to fully consider the changes of all influencing factors. Therefore, after fitting the trend, this paper uses BP neural network model to further analyze the potential missing information in ARIMA model and predict the change of carbon emission intensity more accurately. The specific steps are as follows:

(1) Neural network model selection and structure design.
① Network initialization. According to the input samples, the input dimension, input layer and excitation function are determined.

Set g(x) Equation 11 as Sigmoid function. The expression is as follows:

$$g(x) = \frac{1}{1 + e^{-x}} \quad (11)$$

② output of hidden layer and output layer

BP neural network includes three layers, namely, input layer, output layer and output layer. Let the number of nodes in input layer, hidden layer and output layer be M, N and S. This paper sets five input layers and two hidden layers.

The hidden layer output formula is as follows:

$$H_k = f\left(\sum_{i=1}^{m} x_i \cdot w_{ik} - a_k\right) \quad (12)$$

Where i = 1,2. m, k = 1,2, n, wik is the connection weight coefficient from the input layer to the hidden layer, and ak is the offset.

The output formula Equation 13 of the output layer is:

$$Q_j = \sum_{k=1}^{n} (H_k w_{kj}) - b_j \quad (13)$$

Where j = 1, 2, s; (Equation 12) Is the connection weight coefficient from the hidden layer to the output layer, and is its offset. $w_{kj}$ Is the connection weight coefficient from the hidden layer to the output layer, $b_j$ Is its bias.

③ Calculation error

Error refers Equation 14 to the difference between the output of the input layer and the expected output, and the formula is

$$e_j = \hat{Q} - Q_j \quad (14)$$

Where the expected output value is true. $\hat{Q}$ Represents an expected output value, $Q_j$ For the true value.

④ Update weights

Weight updating formula from input layer to hidden layer (Equations 15, 16):

$$w_{ik} = w_{ik} + \theta H_k \cdot (1 - H_k) \cdot x_i \cdot \sum_{i=1}^{s} w_{ik} e_j \quad (15)$$

$$w_{kj} = w_{kj} + \theta H_k e_j \quad (16)$$

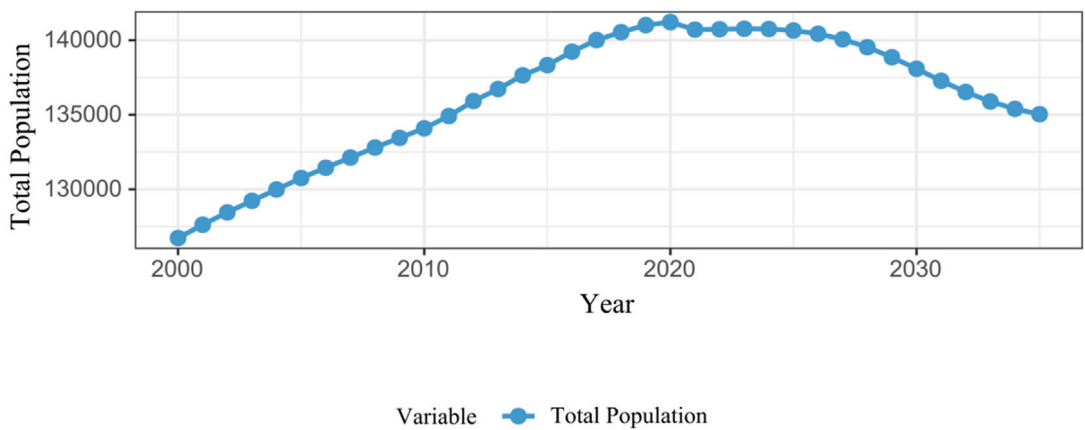

FIGURE 4
Average carbon emissions of provinces from 2000 to 2021 (million tons).

⑤ Update threshold

Threshold updating formula from input layer to hidden layer (Equations 17, 18):

$$a_k = a_k + \theta H_k \cdot (1 - H_k) \cdot x_i \cdot \sum_{j=1}^{s} w_{ik} e_j \quad (17)$$

$$b_j = b_j + e_j \quad (18)$$

⑥ Judge the end condition.

If the error is within a reasonable range, the calculation would finish; otherwise, the hidden layer and the output layer would be recalculated, and the step 2 would be returned backwards.

(2) Division of training and testing data sets

The 21-year data from 2000 to 2021 and the 15-year data from 2022 to 2035 predicted by ARIMA model, including the output value of the three major industries, energy consumption per unit industrial output value and the top 30% of the energy consumption of each industry, are used as the model test set and 70% as the model training set. Carbon emissions are used as the dependent variable, while indicators such as population, economy, and industrial structure are used as independent variables. The collected measured data is used as the training sample set. The training function uses TRAIN-RPROP, the training data is set to 70%, the error function uses ERRORUNC-LNEAR, and the termination function uses STOPFUNC MSE. After iterative operation, the correlation coefficient R is 0.999. The training effect meets expectations, and the accuracy meets the prediction requirements. The model has passed the validity test. The detailed training process curve of BP neural network is shown in Supplementary Appendix SA4.

(3) The fitting effect of 3)BP neural network

In this paper, MSE, RMSE and goodness of fit R are used to evaluate the model effect. The smaller the MSE value, the higher the model accuracy, and the closer R is to 1, the higher the model accuracy.

### 3.2.3 Advantages of ARIMA-BP neural network model

The combination of ARIMA model and BP neural network model is more suitable for carbon emission prediction and analysis in this paper than the single prediction and estimation:

(1) the ability to capture complex patterns. The traditional time series model is sensitive to the quality of data, and requires high quality of data. If there are some problems such as missing values, abnormal values or large noise in the data, the fitting effect and prediction accuracy of the model will be affected. Neural network has a strong ability to capture nonlinear mapping and can capture complex time series patterns and trends. BP neural network can approximate any nonlinear mapping relationship, and the learning algorithm belongs to the global approximation algorithm, which has strong generalization ability, can well solve the shortcomings of ARIMA model, and can better deal with nonlinear mapping relationship and long-term dependence.
(2) Multi-feature processing. Neural network can process multiple features as inputs at the same time, while traditional time series models usually only consider univariate time series. Neural network can make better use of other characteristics related to time series to improve the prediction performance.
(3) Long-term forecasting ability. The recursive structure of neural network (such as long-term and short-term memory network or gated cycle unit) makes it better able to deal with long-term prediction problems. Compared with the traditional time series model, neural network may have better performance in long-term prediction.

## 4 Carbon emission prediction and analysis

### 4.1 Analysis of the overall characteristics of carbon emissions in China

#### 4.1.1 Forecast and analysis of national total carbon emissions

The changing trend of China's overall carbon emissions is shown in Figures 4–6. The total amount of carbon emissions is the sum of the carbon emissions of the three major industries and the carbon emissions of residents' daily consumption. From 2000 to 2035, China's total carbon emissions increased year by year, from 2.575 billion tons to 21.351 billion tons, an increase of 829%. The growth rate of carbon emissions gradually decreased from 4.8% in 2000 to 1.7% in 2035. Before 2018, the growth rate of China's total carbon emissions was relatively slow, and after 2018, the growth rate was relatively fast. The continuous increase of total carbon emissions involves many reasons such as economy, energy and lifestyle, especially the continuous increase of energy demand, which leads to the continuous increase of carbon emissions. At the same time, due to China's adherence to the concept of green development, promoting the synergy of pollution reduction and carbon reduction, clean, low-carbon and efficient utilization of coal, rapid development of new energy sources such as wind energy and solar energy, and vigorous research and development of green and low-carbon technologies, the growth rate of carbon emissions has gradually decreased.

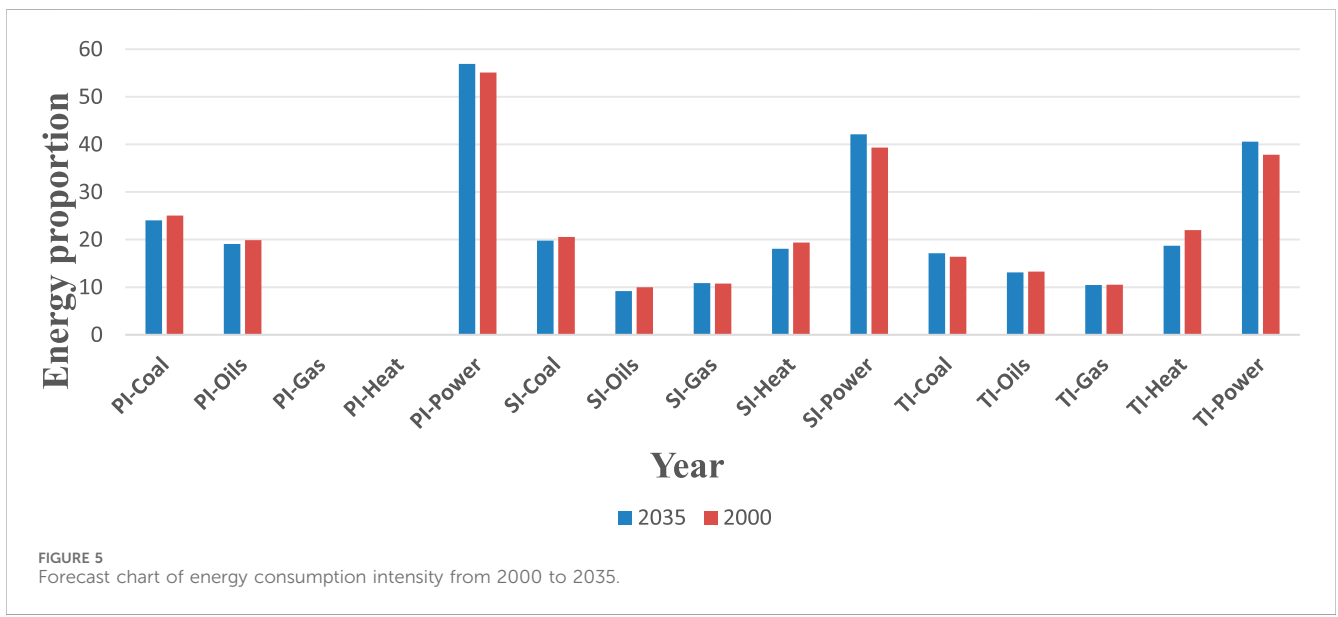

FIGURE 5
Forecast chart of energy consumption intensity from 2000 to 2035.

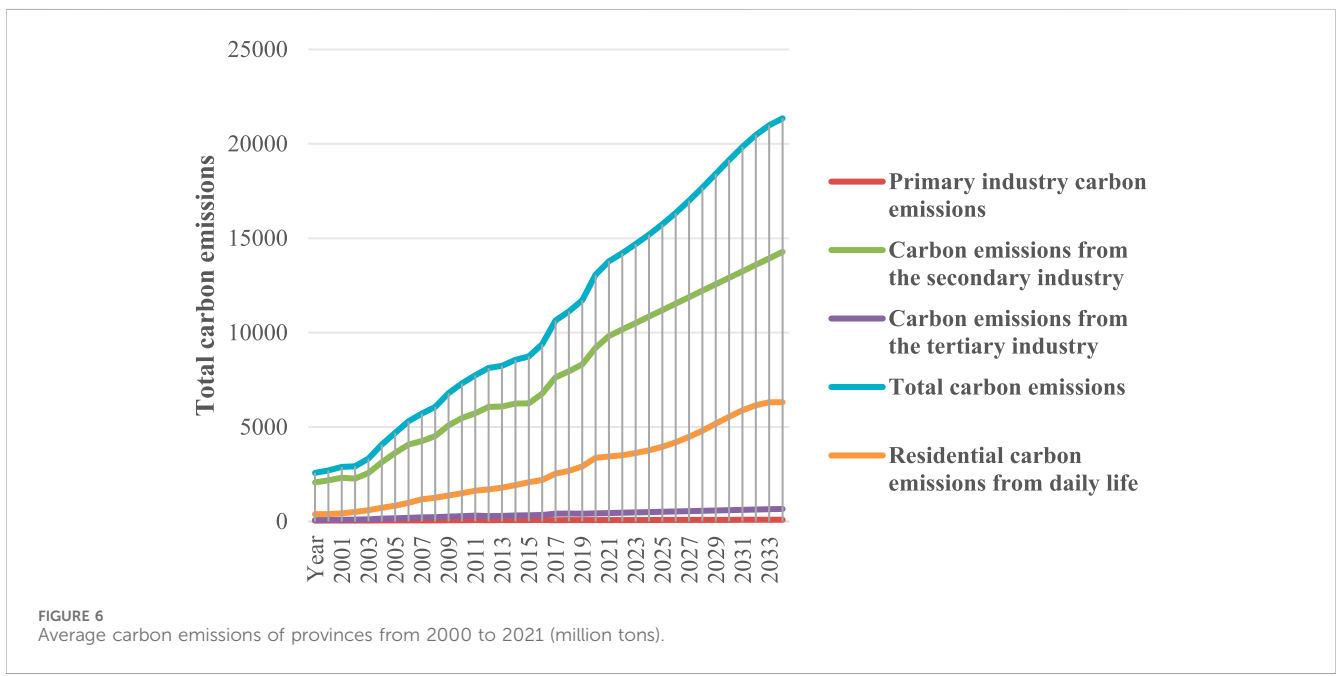

FIGURE 6
Average carbon emissions of provinces from 2000 to 2021 (million tons).

### 4.1.2 Prediction and analysis of carbon emissions of various industries

As can be seen from Figure 6, from 2000 to 2035, the carbon emissions of the primary industry increased from 44 million tons to 92 million tons, with an increase of 208%, the carbon emissions of the secondary industry increased from 2.070 to 14.278 billion tons, with an increase of 690%, and the carbon emissions of the tertiary industry increased from 73 million tons to 663 million tons, with an increase of 91%. Thus, the carbon emissions of the secondary industry are the highest. The economic development has entered a new era and turned to a high-quality development stage. The industrial structure has been further transformed and upgraded, and the proportion of the secondary industry will continue to decline steadily. China is striving to promote the upgrading of manufacturing industry and transform into a high-end manufacturing industry. This will involve the research and development and application of advanced manufacturing technologies, such as artificial intelligence, big data and robotics. With the continuous development of science and technology, agricultural modernization will become the main trend in the future, which will greatly improve the efficiency of agricultural production and enhance the quality and competitiveness of agricultural products. Under the background of a new generation of industrial transformation and upgrading, new urbanization and upgrading of residents' consumption quality, the development of China's tertiary service industry is facing new opportunities, and the leading industry in economic development is further highlighted.

### 4.1.3 Prediction and analysis of carbon emissions of residents' living consumption

As can be seen from Figure 6, from 2000 to 2035, the domestic carbon emissions of Chinese residents increased rapidly, from the lowest 388 million tons to the highest 6.317 billion tons, an increase of 16.42%. With the advancement of urbanization, the total carbon emissions of residents show a significant increase, and urbanization is characterized by low population density expansion, which promotes the continuous growth of carbon emissions. China is in the stage of rapid urbanization, people's requirements for quality of life are gradually improving, the consumption expenditure of various energy goods and services is correspondingly increasing, and the total carbon emissions are constantly increasing. The increase of residents' carbon emissions shows that we should advocate low-carbon life, promote public transportation, encourage walking and cycling, promote green buildings, build energy-saving and environmentally-friendly buildings, reduce building energy consumption, promote consumption upgrading, reduce the total carbon emissions of domestic consumption and promote green development.

## 4.2 Spatial distribution characteristics of carbon emissions in China

As can be seen from Figure 7, the spatial distribution of carbon emissions in China's provinces presents a typical distribution pattern of "East > Central > West" and "North > South", which shows that the regional differences of carbon emissions in China are obvious. Specifically, from 2000 to 2035, the carbon emissions of provinces in China increased continuously. In 2000, the provinces with more carbon emissions were mainly in the eastern region (Figure 8A), In 2000, provinces with higher carbon emissions were mainly located in the eastern region (Figure 8A). In 2021, provinces with higher carbon emissions were mainly located in the eastern and northern region (Figure 8B). This development trend will become more obvious in 2035 (Figure 8C). According to the standard elliptic difference analysis, China's carbon emission center tends to move to the northwest. It shows that while China's power supply and heavy industry moved to the northwest, the processing capacity of terminal treatment facilities failed to keep up in time.

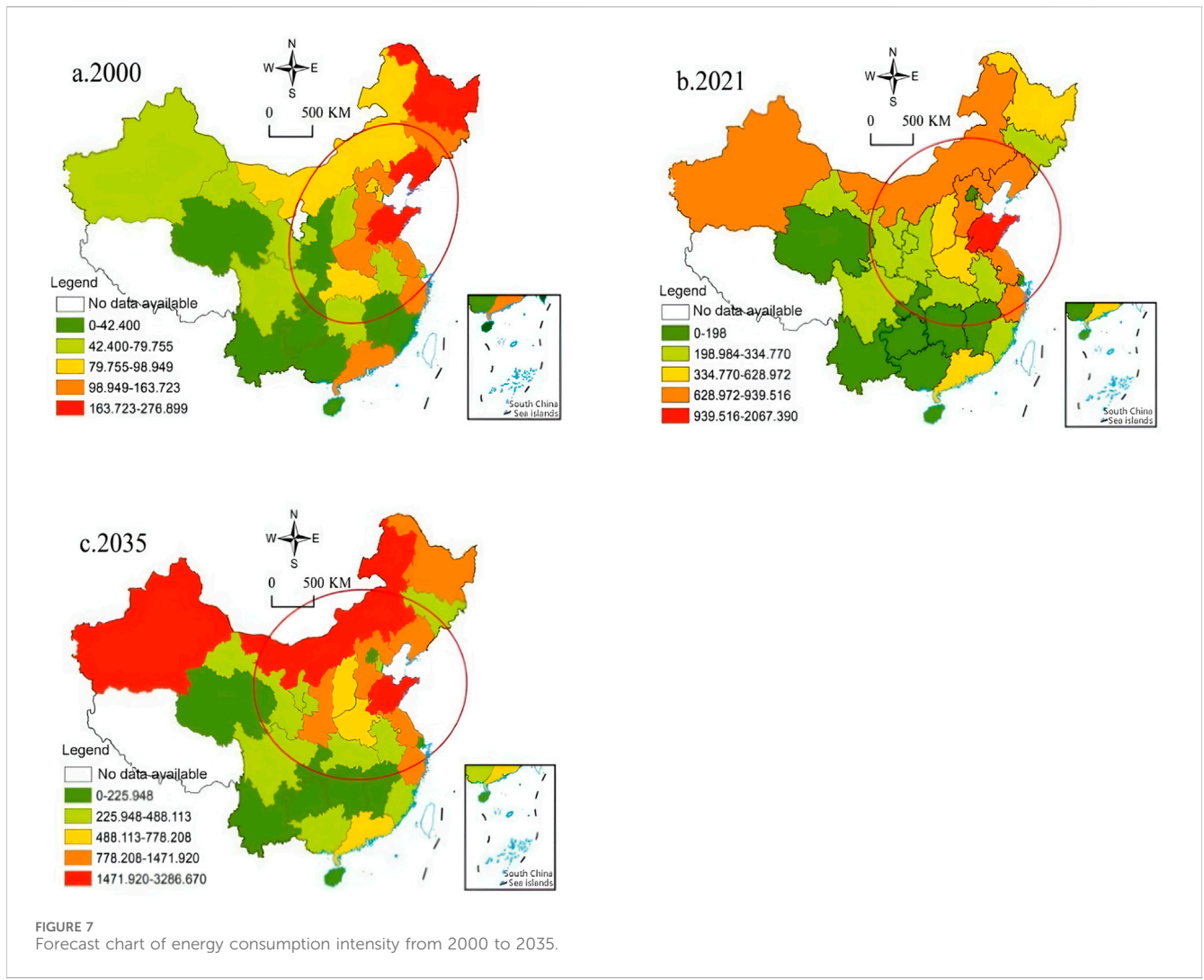

FIGURE 7
Forecast chart of energy consumption intensity from 2000 to 2035.

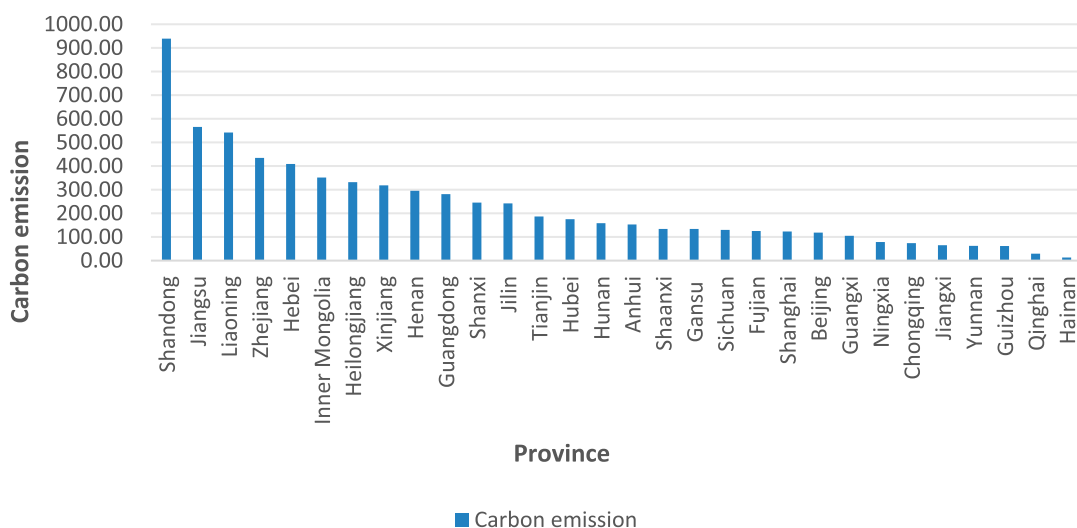

FIGURE 8
Forecast chart of energy proportion from 2000 to 2035.

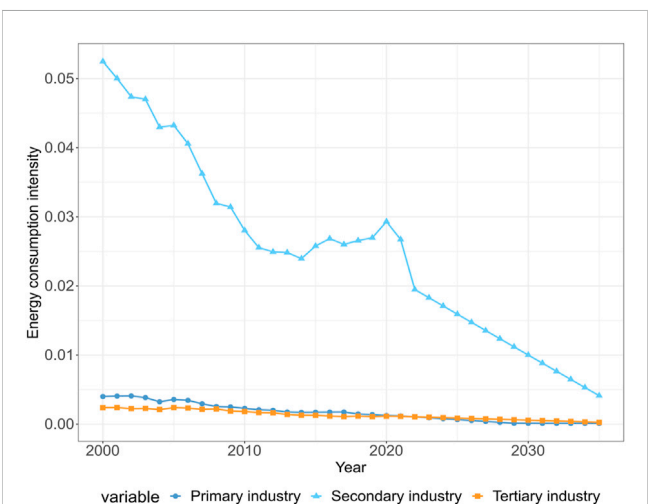

FIGURE 9
Forecast trend of overall carbon emissions in China from 2000 to 2035.

In this paper, the carbon emissions of provinces are further divided into two time periods, 2000-2021 and 2022-2035, and the average carbon emissions of provinces are calculated respectively, and sorted from high to low (Figure 9).

As shown in Figure 10, from 2000 to 2021, Shandong Province had the highest average carbon emissions in China, with an average carbon emission of 940 million tons, and the lowest province is Hainan Province, with an average carbon emission of 12 million tons. The average carbon emission of Shandong Province was 78.25 times that of Hainan Province. Because the industrial structure of Shandong Province is dominated by heavy industries, such as steel, chemical industry, building materials, etc., these industries need a lot of energy in the production process, and the energy utilization efficiency is low, resulting in high carbon dioxide emissions, and the energy consumption structure of Shandong Province is dominated by coal. Although Shandong Province has gradually adjusted its energy structure in recent years and increased the utilization of clean energy such as natural gas and renewable energy, the dominant position of coal consumption is still difficult to change in a short time. Secondly, Shandong Province has a large number of enterprises with high energy consumption and high emission ("two highs"), such as thermal power, steel and chemical industry, which are the main sources of carbon dioxide emissions. Compared with other regions, the carbon emissions of Hainan Island are relatively low, mainly because the economic structure of Hainan Island is dominated by services and agriculture, with relatively few industries, and tourism is the economic pillar. The government of Hainan Island has taken a series of environmental protection measures, such as encouraging the use of clean energy, restricting the entry of highly polluting industries, and carrying out environmental protection activities such as afforestation. In addition, Hainan Island is rich in marine resources, and developing marine economy is also one of the ways to reduce carbon emissions. Nevertheless, Hainan Island, as an important participant in the joint governance of global environmental issues, is still promoting green development, strengthening carbon emission control and mitigating the impact of climate change.

From 2022 to 2035, Shandong Province had the highest average carbon emissions in China, with an average carbon emission of 2.721 billion tons, and the lowest province is Qinghai Province, with an average carbon emission of 93 million tons. Moreover, the average emissions of other provinces have increased compared with 2000–2021, and the trend of change is great. The reason is that China's industrialization and economic growth have led to an increase in energy consumption, especially the use of fossil fuels such as coal and oil. China's population is growing and the process of urbanization is accelerating rapidly. This has led to the increase of buildings, roads and infrastructure, requiring more energy supply. At the same time, traffic volume and industrial activities in cities will also lead to an increase in carbon dioxide emissions. Among them, Jiangsu and Zhejiang have a great decline in the national rankings, while Inner Mongolia and Xinjiang have a great increase. The reason is that Jiangsu and Zhejiang, as developed regions, have high energy utilization efficiency, and high-tech technologies such as photovoltaic power generation and biomass energy have greatly reduced carbon emissions. The rapid economic development in Xinjiang and Inner Mongolia in recent years has led to an increase in energy consumption, especially in manufacturing and electricity and gas sectors, which consume a lot of fossil energy. Low energy utilization efficiency and backward production technology have led to an increase in energy consumption intensity and increased carbon emissions. In the process of urbanization, the growth of urban population, the increase of *per capita* energy consumption and the increase of urban construction investment all consume a lot of energy, thus increasing carbon emissions.

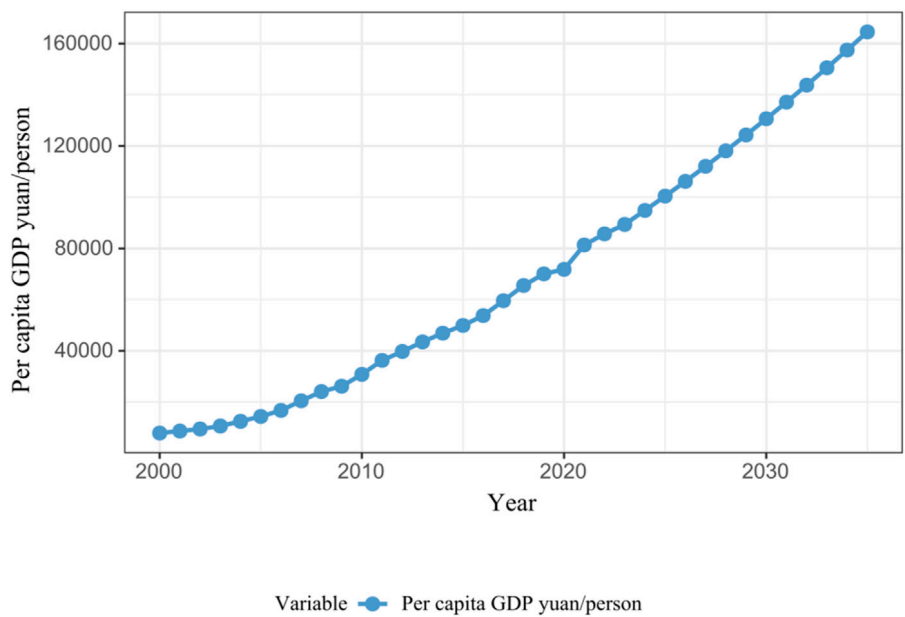

**FIGURE 10**
Forecast of GDP *per capita* from 2000 to 2035 (unit:yuan/person).

## 4.3 Group effect analysis of carbon emissions in China

### 4.3.1 The carbon reduction effect of digital economy

In order to explore the carbon reduction effect of digital economy, the carbon emissions of provinces from 2000 to 2035 were divided into high and low groups according to the development level of digital economy, and the differences were analyzed. Referring to the practice of Zhao et al. (2020), taking the average level of digital economy in 2022 as the standard, 30 provinces in China are divided into provinces with high level of digital economy development and provinces with low level of digital economy development[4]. As can be seen from Table 5, the average carbon emissions of provinces with high level of digital economy development and provinces with low level of digital economy development are 334.127 and 384.774 respectively. Welch's variance test shows that the *p*-value is less than 0.05, which shows that there are significant differences in carbon emissions among provinces with different levels of digital economy development. From the average carbon emission level, the carbon emission of the high digital economy development level group is obviously lower than that of the low digital economy development level group, indicating that the regions with higher digital economy development level have more obvious carbon reduction effect. The main reason is that in areas with high development of digital economy, the carbon reduction effect of digital economy is more obvious due to the comprehensive effect of technological innovation, industrial structure optimization and policy promotion.

---

4 High digital economic development level group for Shanghai, Beijing, Tianjin, Guangdong, Jiangsu, Zhejiang, Hainan, Fujian, shaanxi, low digital economic development level group for Yunnan, Inner Mongolia, Jilin, Sichuan, Ningxia, Anhui, Shandong, Shanxi, Guangxi, Xinjiang, Jiangxi, Hebei, Henan, Hubei, Hunan, Gansu, Guizhou, Liaoning, Chongqing, Qinghai, Heilongjiang.

TABLE 5 Carbon reduction effect of digital economy.

| Group | Sample size | Average value | Standard deviation | Welch's variance test |
|---|---|---|---|---|
| High digital economy | 324 | 334.127 | 318.049 | F = 3.997, $p$ = 0.046** |
| Low digital economy | 756 | 384.774 | 499.073 | |
| amount to | 1,080 | 369.581 | 452.882 | |

TABLE 6 Regional differences of carbon emissions.

| Group | Sample size | Average value | Standard deviation | Welch's variance test |
|---|---|---|---|---|
| Eastern region | 242 | 1.693 | 1.264 | F = 10.354, $p$ = 0.000*** |
| Middle | 176 | 2.150 | 1.569 | |
| The west | 242 | 2.266 | 1.748 | |
| Amount to | 660 | 2.025 | 1.556 | |

TABLE 7 Industrial structure differences of carbon emissions.

| Group | Sample size | Average value | Standard deviation | Variance test |
|---|---|---|---|---|
| Low | 441 | 2.196 | 1.588 | F = 16.388, $p$ = 0.000*** |
| Tall | 219 | 1.681 | 1.432 | |
| Amount to | 660 | 2.025 | 1.556 | |

### 4.3.2 Regional differences in carbon emissions

This paper classifies the provinces into three major regions: east, central, and west. The differences of regional carbon emission intensity are analyzed. As can be seen from Table 6, there are significant differences in carbon emission intensity in different regions. The average carbon emission intensity in the eastern region is 1.69%, which is the lowest among the three regions. The economic development in the eastern region is the most advanced, and the industrial structure is more environmental-friendly. High-tech industries, service industries and other low-carbon industries account for a relatively high proportion. These industries usually have high energy utilization efficiency, so the carbon emission intensity is low. The average carbon emission intensity in the central region is 2.15%, which is slightly higher than that in the eastern region. The central region is in the stage of rapid industrialization, and these industries usually need to consume a lot of fossil energy in the production process, resulting in relatively high carbon emission intensity. With the gradual optimization and transformation and upgrading of the economic structure in the central region, the intensity of carbon emissions has gradually decreased. The average carbon emission intensity in the western region is 2.27%, which is the highest among the three regions. This is related to the relatively backward economic development, insufficient optimization of industrial structure and more dependence on traditional high-carbon industries, such as coal and petrochemical industry.

### 4.3.3 Industrial structure differences of carbon emissions

This paper refers to Gan et al. (2011) and uses the ratio of the output value of the tertiary industry to the secondary industry as the indicator of industrial structure, and divides the provinces into two groups based on the average value. From Table 7, it can be seen that the average carbon emission intensity of the high and low groups is 1.681 and 2.196 respectively, indicating that there are significant differences in carbon emission intensity among provinces with different development levels of industrial structure. When the level of industrial structure is low, it may rely more on high-carbon industries such as traditional manufacturing, which usually consume a lot of fossil energy in the production process, resulting in high carbon emission intensity. In areas with a high level of industrial structure, they more rely on high-tech industries, service industries and other low-carbon industries. These industries usually have low demand for energy, high utilization efficiency, and tend to use clean energy technologies, so the carbon emission intensity is relatively low.

### 4.3.4 New qualitative productivity differences of carbon emissions

Referring to the research of Jia et al. (2024), this paper creates a new quality productivity index system, and divides the new quality productivity development levels of each province over the years into two groups based on the average level. From Table 8, it can be seen that there are significant differences in carbon emission intensity in different regions, and the average carbon emission intensity of high and low groups is 1.069 and 1.832, indicating that the carbon emission intensity is relatively low in regions with higher development level of new quality productivity. New-quality productivity refers to the advanced productivity that innovation plays a leading role. Therefore, the difference of carbon emissions is mainly due to the influence of four factors: the intensity of technological innovation, the degree of optimal allocation of production factors, the upgrading of industrial structure and policy orientation. Areas with high development level of new quality productivity can effectively reduce the intensity of carbon emissions and realize green and low-carbon development.

TABLE 8 Difference of new quality productivity of carbon emissions.

| Group | Sample size | Average value | Standard deviation | Welch's variance test |
|---|---|---|---|---|
| Tall | 98 | 1.069 | 0.731 | $F = 69.910, p = 0.000^{***}$ |
| Low | 262 | 1.832 | 1.358 | |
| Amount to | 660 | 2.025 | 1.556 | |

TABLE 9 Prediction results of carbon emission intensity of three major industries.

| Industry | Collect | MSE | RMSE | MAE | MAPE | R2 |
|---|---|---|---|---|---|---|
| Primary industry | Training set | 0.002 | 0.039 | 0.028 | 6.221 | 0.998 |
| | Test set | 0 | 0.004 | 0.004 | 0.969 | 0.912 |
| Secondary industry | Training set | 0.097 | 0.312 | 0.265 | 13.485 | 0.932 |
| | Test set | 0.633 | 0.795 | 0.783 | 39.613 | 0.916 |
| Service sector | Training set | 0.002 | 0.047 | 0.04 | 12.999 | 0.999 |
| | Test set | 0.001 | 0.036 | 0.036 | 11.437 | 0.998 |

# 5 Analysis of driving factors of carbon emissions

## 5.1 Prediction of driving factors of carbon emissions

### 5.1.1 Prediction of energy consumption intensity

As can be seen from Table 9, the MSE fluctuation range of carbon emission intensity of the three major industries is 0–0.633, and the regression coefficient is high and close to 1, which shows that the simulation prediction effect is significant and the prediction data is reliable. As can be seen from Figure 7, from 2000 to 2035, China's industrial energy consumption intensity was "secondary industry > tertiary industry > primary industry", the energy intensity of primary industry decreased from 0.0039 to 0.0001, while the secondary industry decreased from 0.0525 to 0.0041, and the tertiary industry decreased from 0.0024 to 0.0003. It shows that the reduction of China's energy consumption intensity is mainly concentrated in the secondary industry. Although there is still a certain gap between China's energy intensity and that of European and American countries, with the adjustment of China's industrial structure, diversification of energy structure and continuous improvement of industrialization level, there is still room for a certain decline in energy intensity.

### 5.1.2 Energy proportion forecast

The ratio of energy consumption to total energy consumption in an industry reflects the energy use, energy consumption structure and industrial structure changes, and plays an important role in determining carbon emissions. As can be seen from Table 9, from 2010 to 2035, the proportion of coal energy use in China's primary industry decreased, while the proportion of oil energy use and electric energy use increased. It shows that China has experienced the transformation from coal-based energy structure to diversified energy structure. The use of electricity energy and natural gas energy in the secondary industry has increased, while the use of coal, oil products and thermal energy has decreased. The use of thermal energy, coal energy and oil energy in the tertiary industry decreased, while the use of natural gas energy and electric energy increased. With the increase of the proportion of clean energy, the carbon emissions of traditional energy consumption are also decreasing.

TABLE 10 Error table of population size forecast.

| | MSE | RMSE | MAE | MAPE | $R^2$ |
|---|---|---|---|---|---|
| Training set | 6,885.4502 | 262.401 | 210.971 | 2.494 | 0.99 |
| Test set | 42,387.0595 | 651.053 | 640.503 | 7.522 | 0.98 |

TABLE 11 Per capita GDP error table.

| | MSE | RMSE | MAE | MAPE | $R^2$ |
|---|---|---|---|---|---|
| Training set | 756.146 | 2,729.465 | 2,356.524 | 55.343 | 0.998 |
| Test set | 815.624 | 9,065.259 | 8,872.062 | 203.335 | 0.997 |

### 5.1.3 Population size and economic forecast

From Tables 10, 11, it can be seen that the average MSE of China's population size forecast is 24.63, and the population forecast regression coefficient R = 1, the average MSE of *per capita* GDP forecast is 0.07855, and the *per capita* GDP forecast regression coefficient R2 = 0.99. It shows that the prediction effect of the model is good. Assume that the GDP growth rate is constant at 4% from 2024 to 2035. As can be seen from Figure 12, from 2000 to 2023, the population size gradually increased, from 1,267.43 million in 2000 to 1,407.69 million in 2023, and gradually decreased from 2024 to 2035. China's population size reached its maximum in 2023. Due to economic pressure, living costs and other pressures, Chinese people's fertility rate will decline, the aging population and other factors would lead to the gradual reduction of China's population size.

TABLE 12 LMDI annual decomposition effect.

| Year | Energy use Structural effects | Energy use Strength effect | Per capita GDP effect | Population size effect | Industrial structure effect | Gross effect; ensemble |
|---|---|---|---|---|---|---|
| 2024 | −1,531.48 | 2,861.17 | −2,420.52 | −46.08 | 36.55 | −1,100.36 |
| 2025 | −1,659.07 | 3,043.93 | −2,596.87 | −51.13 | 40.29 | −1,222.86 |
| 2026 | −1786.55 | 3,226.85 | −2,773.05 | −56.18 | 44.03 | −1,344.91 |
| 2027 | −1913.91 | 3,409.91 | −2,949.07 | −61.21 | 47.76 | −1,466.52 |
| 2028 | −2041.15 | 3,593.13 | −3,124.93 | −66.24 | 51.50 | −1,587.69 |
| 2029 | −2,168.26 | 3,776.49 | −3,300.62 | −71.27 | 55.25 | −1708.41 |
| 2030 | −2,295.26 | 3,960.01 | −3,476.14 | −76.28 | 58.99 | −1828.68 |
| 2031 | −2,422.14 | 4,143.68 | −3,651.50 | −81.29 | 62.74 | −1,948.52 |
| 2032 | −2,548.90 | 4,327.50 | −3,826.70 | −86.30 | 66.48 | −2067.90 |
| 2033 | −2,675.54 | 4,511.48 | −4,001.73 | −91.29 | 70.23 | −2,186.85 |
| 2034 | −2,802.06 | 4,695.60 | −4,176.60 | −96.28 | 73.98 | −2,305.35 |
| 2035 | −2,928.46 | 4,879.88 | −4,351.30 | −101.27 | 77.74 | −2,423.41 |

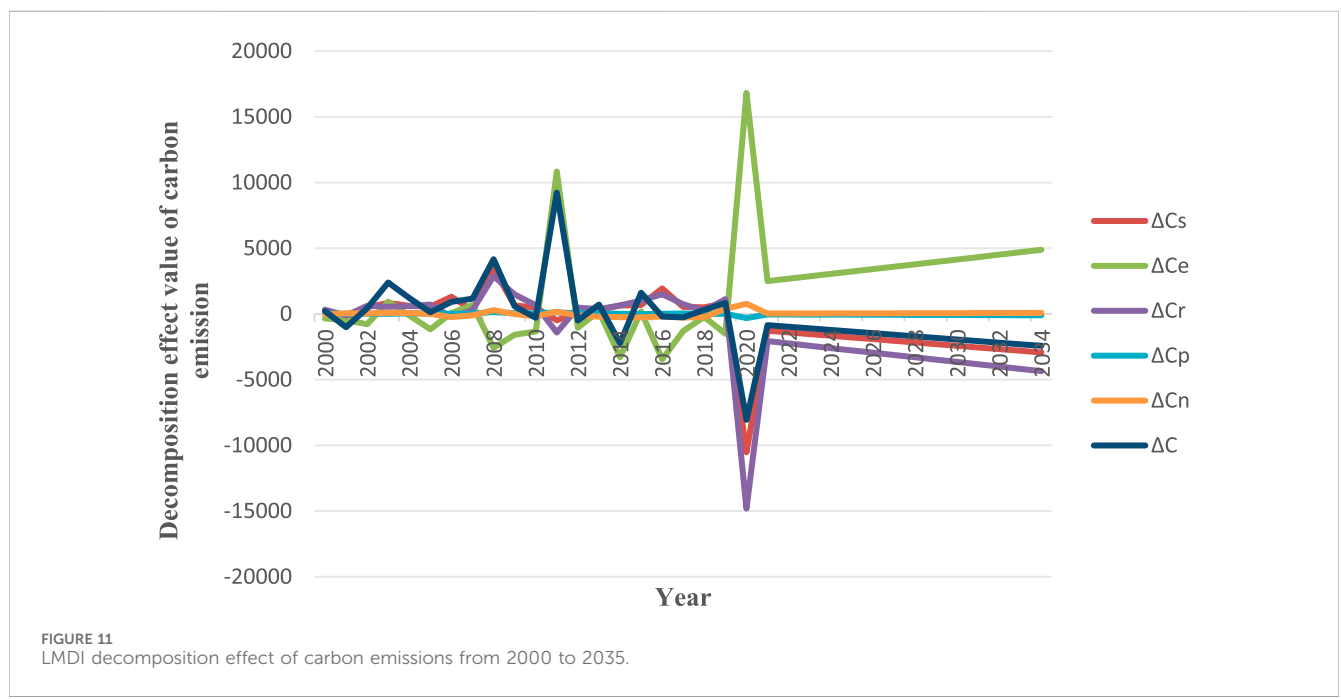

FIGURE 11
LMDI decomposition effect of carbon emissions from 2000 to 2035.

As can be seen from Table 12, from 2000 to 2035, due to the combined effects of economic development, the promotion of scientific and technological innovation and the stable social environment, the *per capita* GDP showed a growing trend, from 7,942 yuan in 2000 to 164,556 yuan in 2035. From 2000 to 2035, the growth rate of *per capita* GDP increased from 9.7% to 11.7% and then to 4.5%. Due to the impact of the COVID-19 epidemic, the *per capita* GDP growth rate dropped to 5.3% in that year. Since then,

China's economy has recovered rapidly and the *per capita* GDP growth rate has reached 5.9%. Since then, the *per capita* GDP growth rate has gradually increased and stabilized.

## 5.2 Decomposition of LMDI carbon emission factors

Using LMDI decomposition method to decompose the carbon emission factors in China from 2000 to 2035 into four factors: the annual and cumulative effects of energy intensity, energy structure, industrial structure, economic development and population size.

They are shown in Figure 11 and Table 12. Equations 5–10 to calculate the carbon emission driven decomposition effect value.

From the year-by-year decomposition results in Table 12, we can see that from 2000 to 2035, the energy consumption intensity effect varies greatly, and in most years it promotes the growth of carbon emissions, with the highest value of 16,816.83, making the largest contribution. With the adjustment of China's industrial structure, the diversification of energy structure and the continuous improvement of industrialization level, there is still room for the decline of energy consumption intensity. The structural effect of energy consumption shows a fluctuating trend, which has a relatively small impact on the growth of carbon emissions. It can promote carbon emissions before 2020, and then it mainly shows an inhibitory effect, with the highest value of −10528.96. The effect of industrial structure changes little year by year, and it has little influence on the growth of carbon emissions. Before 2019, it mainly inhibits carbon emissions, and the following years mainly promote it. Per capita GDP mainly inhibits the growth of carbon emissions, with the highest value of-14,812.49, and the contribution rate is high. The population size effect has little influence on carbon emissions. Since the reform and opening up, China's economy has developed by leaps and bounds, and the secondary industry has become the main driving force of economic growth. At the same time, economic development requires a lot of energy consumption, which has also caused the continuous increase of carbon emissions. In the future, by upgrading the technical level and optimizing the industrial structure, carbon emissions will gradually slow down with the economic development.

# 6 Conclusions and recommendations

## 6.1 Research conclusions

(1) The ARIMA-BP neural network model is used to predict the total carbon emissions of provinces from 2000 to 2035. Overall, China's total carbon emissions are increasing year by year, but with the continuous optimization of industrial structure, the growth rate of carbon emissions is gradually decreasing. From the specific structural changes of carbon emissions, the total amount of carbon emissions is "secondary industry > residents' life > tertiary industry > primary industry", in which the growth rate of carbon in secondary industry and residents' life is faster, while the change trend of primary industry and tertiary industry is smaller.

(2) Using ArcGIS to describe the spatial distribution characteristics of carbon emissions in China's provinces. The spatial distribution of carbon emissions in each province presents a typical unbalanced distribution pattern of "eastern > central > western" and "northern > southern", which reflects the characteristics of carbon emissions in different stages of economic development. According to the standard elliptic difference analysis, the carbon emission center will move to the northwest from 2000 to 2035.

(3) Based on the new development concept, the carbon emissions of each province are divided into different groups, and the heterogeneity of group effect is analyzed. The carbon emissions in areas with high level of digital economy, advanced industrial structure and new quality productivity are relatively small, which has significant group difference effect, indicating the importance of industrial structure and new formats in emission reduction.

(4) Based on LMDI decomposition method, the driving factors of energy carbon emissions from 2000 to 2035 are analyzed. China's energy consumption intensity is "secondary industry > tertiary industry > primary industry", and the reduction of carbon emission intensity is mainly concentrated in the secondary industry, and the energy structure has gradually changed from coal-based to diversified energy structure; The intensity effect of energy consumption is the main factor driving the continuous growth of carbon emissions, and the *per capita* GDP and energy consumption structure effect are the main factors restraining carbon emissions. The population size effect has a promoting effect on the growth of carbon emissions in the first year, but most of the later years have a restraining effect, and the industrial structure effect has a relatively small impact.

## 6.2 Policy recommendations

### 6.2.1 Optimize the energy consumption structure

Through this study, it was found that as the proportion of clean energy utilization increases, the total carbon emissions will correspondingly decrease. Therefore, it is necessary to optimize the energy structure. First, government should further improve the regulation of total energy consumption and intensity, focus on controlling fossil energy consumption, optimize energy consumption structure, and gradually shift to the "dual control" system of total carbon emission and intensity. Second, government should coordinate industrial restructuring, pollution control, ecological protection, and response to climate change. By optimizing the energy system and popularizing energy-saving technologies and equipment, we will vigorously develop clean energy, promote clean energy represented by hydropower and nuclear power, reduce energy consumption, and jointly promote carbon reduction, pollution reduction, greening and growth. Third, government should effectively implement the policy of giving priority to energy conservation, make intensive use of energy resources throughout the whole process, all fields and all links of economic and social development, establish the concept of thrift and thrift, and accelerate the formation of an energy-saving society.

### 6.2.2 Promote the upgrading of industrial structure

According to historical data and forecast results, China's secondary industry has a high carbon emission intensity and volume. Fully developing the tertiary industry can help optimize the industrial structure and reduce carbon emissions. First, government should promote agricultural modernization, popularize green agricultural technology, encourage farmers to adopt conservation tillage, precision agricultural technology and biotechnology, improve the efficiency and quality of agricultural production, and reduce carbon emissions. It is necessary to optimize the structure of agricultural industry, rationally adjust the structure of planting and aquaculture, and promote the transformation and upgrading of agricultural industry. Secondly, government should promote the high-end manufacturing industry, develop green manufacturing technology, encourage enterprises to adopt advanced production technology and equipment, improve production efficiency and product quality, and reduce energy consumption and carbon emissions. Policymaker should optimize the industrial structure, accelerate the elimination of backward production capacity and high-pollution industries, develop emerging industries and green industries, and improve the green and low-carbon level of manufacturing. Finally, government should develop the tertiary industry, develop the green service industry and strengthen the energy management of the service industry.

### 6.2.3 Develop new quality productivity

The carbon emission intensity is low in areas with high development level of new quality productivity. To achieve green and low-carbon development, new quality productivity is the key to reduce carbon emissions. By promoting innovative technologies and sustainable development, we will achieve more efficient and cleaner production methods to reduce carbon emissions; Improve production efficiency through scientific and technological innovation, such as actively cultivating strategic emerging industries such as new energy, new materials, advanced manufacturing and electronic information, reducing China's resource consumption and environmental pollution, and realizing sustainable development; Based on the regional imbalance of carbon emissions, establish and improve the carbon trading market to provide a platform for carbon emission trading. The government should formulate relevant policies, clarify the rules and standards of carbon trading, provide legal protection for carbon trading, and strengthen the supervision of carbon trading market. Promote provincial carbon emission control policies, targeting heavy carbon emitting industries such as metallurgy, chemical industry, light industry, and building materials in Shandong and other provinces, strengthen the regulation of steel production capacity and output, deeply adjust the structure of steel products, and accelerate the energy-saving and carbon reduction transformation of the steel industry. Continuously improving the ecological carbon sequestration capacity of provinces such as Inner Mongolia and Jilin. We will comprehensively promote the management of the mountain, water, forest, farmland, lake, grass and sand system, continue to implement the protection of protective forests and natural forests such as the Three North and Yangtze River, construct high standard farmland, protect and restore wetlands, return farmland to forests and grasslands, restore grassland ecology, and comprehensively control desertification and rocky desertification.[5]

### 6.2.4 Promote digital transformation

The carbon emission intensity is low in areas with high level of digital economy development. Developing digital economy is an important way to reduce carbon emissions. Digital transformation can reduce dependence on traditional industries and reduce energy consumption and carbon emissions. Optimize supply chain management, reduce energy consumption, build digital infrastructure, such as building green data centers, promoting the use of energy-saving equipment and technologies, and improving the energy efficiency of data centers. At the same time, use renewable energy to power the data center and reduce carbon emissions. Promote 5G and Internet of Things technologies, and realize the interconnection between devices through 5G and Internet of Things technologies, improve energy utilization efficiency, and reduce unnecessary energy consumption and carbon emissions.

## Data availability statement

The original contributions presented in the study are included in the article/Supplementary Material, further inquiries can be directed to the corresponding authors.

## Author contributions

SZ: Conceptualization, Data curation, Formal analysis, Software, Writing–original draft, Visualization, Writing–review and editing. ZL: Project administration, Resources, Writing review and editing. HD: Funding acquisition, Investigation, Writing review and editing. XY: Methodology, Writing review and editing. JT: Writing–review and editing. BY: Validation, Writing review and editing. ZZ: Supervision, Writing–review and editing.

## Funding

The author(s) declare that financial support was received for the research, authorship, and/or publication of this article. General Project of Hunan Provincial Social Science Foundation (23YBA289). Undergraduate innovation and Entrepreneurship Program (Hunan University of Finance and Economics 2024064).

## Conflict of interest

The authors declare that the research was conducted in the absence of any commercial or financial relationships that could be construed as a potential conflict of interest.

## Publisher's note

All claims expressed in this article are solely those of the authors and do not necessarily represent those of their affiliated organizations, or those of the publisher, the editors and the reviewers. Any product that may be evaluated in this article, or claim that may be made by its manufacturer, is not guaranteed or endorsed by the publisher.

## Supplementary material

The Supplementary Material for this article can be found online at: https://www.frontiersin.org/articles/10.3389/fenvs.2024.1497941/full#supplementary-material

---

5 White Paper on China's Policies and Actions to Address Climate Change 2021(https://www.gov.cn/zhengce/2021-10-27/content_5646697.htm).

# References

Chen, C., He, Yu, Xuan, C., Zihan, Q., and Songtao, L. (2024). *Carbon emission scenario prediction and emission reduction potential analysis of power grid enterprises based on LEAP model [J/OL]*. Journal of North China Electric Power University Natural Science Edition, 1–8.

Cheng, S. (2023). *Research on energy demand and carbon emissions in Jilin province based on LEAP model [D]*. Jilin University.

Gan, C., Zheng, R., and Yu, D. (2011). The influence of industrial structure change on economic growth and fluctuation in China. *Econ. Res.* 46 (05), 4–16+31.

Hu, J., Luo, Z., and Li, F. (2022). Prediction of China's carbon emission intensity under the goal of "peak carbon dioxide emissions"-analysis based on LSTM and ARIMA-BP model. *Financial Sci.* (02), 89–101.

Hu, Z., Chen, C., and Baojian, Hu (2021). Remote sensing inversion of PM2.5 concentration in China based on spatio-temporal XGBoost. *J. Environ. Sci.* 41 (10), 4228–4237.

Jia, S., Zhang, J., and Yi, P. (2024). Research on the influence of ESG development on the new quality productivity of enterprises-empirical evidence from China A-share listed enterprises [J/OL]. *Contemp. Econ. Manag.* 1-13.

Liu, Z. (2018). *Study on calculation, convergence and decoupling of carbon emissions from regional energy consumption in China [M]*. Beijing: China Financial and Economic Press, 29–43.

Lu, Y., Xuan, W., and Liwei, Z. (2024). Spatial-temporal pattern evolution of carbon emissions in Anhui Province and peak carbon dioxide emissions path prediction-based on STIRPAT expansion model and ridge regression model. *Regional Res. Dev.* 43 (01), 146–152+173.

Nan, W., Zhanshuo, H., Wang, L., Lv, X., and Zhang, Y. (2023). Comprehensive application practice analysis of exponential smoothing method model and ARIMA model in predicting industry carbon emission trends. *Digital Technol. Appl.* 41 (12), 91–93. doi:10.19695/j.cnki.cn12-1369.2023.12.29

Ni, L., Gu, W, Zhou, T., Hao, P., and Jiang, J. (2025). Leak aperture recognition of natural gas pipeline based on variational mode decomposition and mutual information. *Measurement* 242 (PC), 116017. doi:10.1016/j.measurement.2024.116017

Wang, Y. (2005). *Applied time series analysis [M]*. Beijing: Renmin University of China Press.

Xu, Y., Huang, X., Zheng, X., Zeng, Z., and Jin, T. (2024). VMD-ATT-LSTM electricity price prediction based on grey wolf optimization algorithm in electricity markets considering renewable energy. *Renewable Energy* 236, 121408. doi:10.1016/j.renene.2024.121408

Yuli, S., Dabo, G., Heran, Z., Li, Y., Ou, J., Meng, J., et al. (2018). China CO2 emission accounts 1997–2015. *Sci. data* 5 (1), 170201. doi:10.1038/sdata.2017.201

Zhao, T., Zhang, Z., and Shangkun, L. (2020). Digital economy, entrepreneurial activity and high-quality development-empirical evidence from China. *Manag. World* 36 (10), 65–76.

Zhou, D., Ma, T., Li, Q., and Zhu, B. (2024). Research on the influencing factors of carbon emissions and carbon peak prediction in Jiangsu Province based on STIRPAT model - taking the industrial sector as an example. *Business Exhib. Econ.* (06), 146–150. doi:10.19995/j.cnkiCN10-1617/F7.2024.06.146

Zhu, L., Dabo, G., Wei, W., Davis, S. J., Ciais, P., Bai, J., et al. (2015). Reduced carbon emission estimates from fossil fuel combustion and cement production in China. *Nature* 524 (7565), 335–338. doi:10.1038/nature14677